\documentclass[12pt]{article} 

\usepackage[utf8]{inputenc} 
\usepackage{amsmath,amsthm,verbatim,amssymb,amsfonts,amscd,graphicx}
\usepackage[colorlinks,citecolor=blue,urlcolor=blue]{hyperref}

\usepackage{geometry} 
\usepackage{graphicx} 

\usepackage{dsfont}

\usepackage{booktabs} 
\usepackage{array} 
\usepackage{verbatim} 
\usepackage{subfig} 
\usepackage{amsmath}
\usepackage[utf8]{inputenc}
\usepackage[english]{babel}

\usepackage{algorithm}
\usepackage{algorithmic}
\usepackage{bm}

\usepackage[symbol]{footmisc}

\usepackage{fancyhdr} 
\usepackage{sectsty}
\allsectionsfont{\sffamily\mdseries\upshape} 

\usepackage[nottoc,notlof,notlot]{tocbibind} 
\usepackage{natbib}
\usepackage{bbm}
\usepackage[titles,subfigure]{tocloft} 
\usepackage{comment}
\usepackage{stmaryrd} 
\usepackage{enumitem} 

\usepackage{xr}
\newcommand{\p}{{\rm I}\kern-0.18em{\rm P}}
\newcommand{\1}{{\rm 1}\kern-0.24em{\rm I}}
\newcommand{\E}{{\rm I}\kern-0.18em{\rm E}}
\newcommand{\R}{{\rm I}\kern-0.18em{\rm R}}

\newcommand{\mC}{\mathcal{C}}

\title{Response to Comment by Schilling}
\date{}
\author{\textsc{Jay Bartroff}$^*$, \textsc{Gary Lorden}$^\dag$, and \textsc{Lijia Wang}$^\ddag$\\
\small{$^*$Department of Statistics and Data Sciences, University of Texas at Austin, Austin, Texas, USA}\\
\small{$^\dag$Department of Mathematics, Caltech, Pasadena, California, USA}\\
\small{$^\ddag$Department of Mathematics, University of Southern California, Los Angeles, California, USA}
}

\begin{document}
\maketitle

We appreciate the recent paper of \citet[][hereafter SS]{Schilling22} on confidence intervals for the hypergeometric being brought to our attention, which we were not aware of while preparing our paper~(Bartroff, Lorden, and Wang, \citeyear{Bartroff22}, hereafter BLW) on that subject.   Although there are commonalities between the two approaches, there are some important distinctions that we highlight here.  Following those papers' notations, below we denote  the confidence intervals for the hypergeometric success parameter based on sample size~$n$ and population size~$N$ by LCO for SS, and $\mC^*$  for  BLW. In the numerical examples below, LCO (\url{github.com/mfschilling/HGCIs}) and $\mC^*$  (\url{github.com/bartroff792/hyper}) were computed using the respective authors' publicly available R code, running on the same computer.

\begin{figure}[!ht]
    \centering
    \caption{The computational time of the confidence intervals $\mC^*$ and LCO for $\alpha=.05$, $N=200, 400, \ldots, 1000$, and $n=N/2$.} \label{time.LCO.us.1000}
\includegraphics[width=12cm]{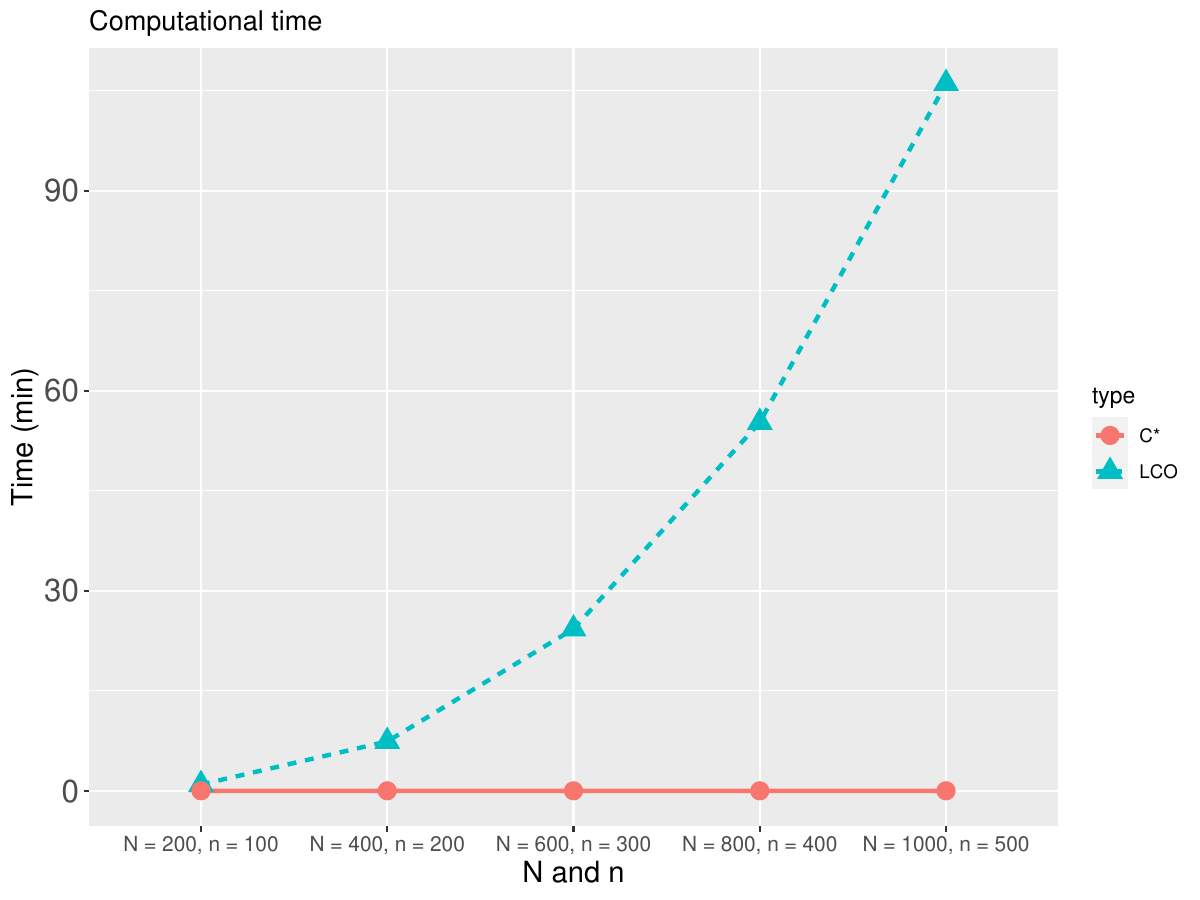} 
\end{figure}

\textbf{Computational time.} LCO and $\mC^*$ differ drastically in the amount of time required to compute them. Figure~\ref{time.LCO.us.1000} shows the computational time of LCO and $\mC^*$ for $\alpha = 0.05$, $N=200, 400, \ldots, 1000$, and $n=N/2$. For example, for  $N = 1000$ the computational time of LCO exceeds $100$ minutes whereas $\mC^*$ requires roughly $1/10$th of a second ($0.002$ minutes).   In further numerical comparisons not included here, we found this relationship to be common for moderate to large values of the sample and population sizes, $n$ and $N$. This may be due to the algorithm for computing LCO  which calls for searching \textit{``among all acceptance functions of minimal span''} (SS, p.~37).

\textbf{Provable optimality.}  SS contains two proofs, one in the Appendix of a basic result about the hypergeometric parameters, and one in the main text of the paper's only theorem (SS, p.~33) which is  a general result that size-optimal hypergeometric acceptance sets are inverted to yield size-optimal confidence ``intervals.''  However, not all inverted acceptance sets will yield proper intervals, and in practice one often ends up with non-interval confidence \emph{sets}, e.g., intervals with ``gaps.'' This occurs when the endpoint sequences of the acceptance intervals being inverted are non-monotonic, or themselves have gaps.  SS address this  by modifying their proposal in this situation to mimic a method of \citet{Schilling14} developed for the Binomial distribution.   SS (p.~36--37) write, \textit{``Where there is a need to resolve a gap, in which case the minimal span acceptance function that causes the gap is replaced with the one having the next highest coverage.''}  This modification is nontrivial in that it, in general, changes both the length and coverage probability of the LCO acceptance and confidence intervals.  Since their optimality argument relies on choosing the acceptance functions with the \emph{highest} coverage, does size optimality of LCO still hold when instead choosing the ``next highest'' coverage, and utilizing  a technique for the Binomial that has not been verified for the hypergeometric?  These questions are not addressed mathematically in SS, and similar questions remain about their proposed confidence intervals for the population size~$N$. On the other hand, BLW develops a complete optimality theory for $\mC^*$, which is substantial (requiring 21 theorems and lemmas in the paper and its supplement) and includes sufficient conditions for when gaps in the optimal confidence sets make $\mC^*$ sub-optimal. This is rare and, even when it happens, only causes $\mC^*$ to be ``too big'' by at most a single point.

\textbf{Symmetry.}  In hypergeometric inference, any procedure being used should give the same result whether the binary property being counted is considered ``success'' or ``failure,''  such distinctions being arbitrary. BLW call this property ``symmetry'' while SS  call it ``equivariance'' and write, \textit{``Equivariance is appropriate when estimating the success parameter of the hypergeometric distribution''} (SS, p.~35).  The $\mC^*$ intervals are symmetrical  by construction and their optimality theory in BLW takes this into account, which is necessarily more complex because of it. When considering optimality, achieving symmetry is nontrivial and is not just a matter of, say, replacing an asymmetric interval by the reflection of its counterpart since this could  change both the interval's length and coverage probability.  On the other hand, the LCO intervals are asymmetric by design, and asymmetries occur frequently.  In the same setting as Figure~\ref{time.LCO.us.1000},  Table~\ref{violate.LCO.us.1000} contains the proportion of LCO's upper and lower endpoint pairs that are asymmetric.

\begin{table}[!ht]
\caption{The proportion (``\% Asymmetric,'' to 1 decimal place) of LCO upper and lower endpoint pairs that are asymmetric
for $\alpha=.05$, $N=200, 400, \ldots, 1000$, and $n=N/2$.}\label{violate.LCO.us.1000}
\begin{center}
\begin{tabular}{c|c|c|c|c|c}
\hline
$N$ & 200 & 400 & 600 & 800 & 1000\\\hline
\% Asymmetric &   80.2\% & 76.1\%& 73.1\% & 77.1\% & 73.5\%\\\hline
\end{tabular}
\end{center}
\label{table:air}
\end{table}

Because of these distinctions between the SS and BLW methods, we hope that readers will regard the two papers as distinct though partially overlapping contributions to the statistics
literature, and will consider the optimality results proved in BLW as supporting many of the methodological recommendations
in both papers.

\bigskip

\noindent\textbf{Disclosure:} The authors report there are no competing interests to declare.


\begin{thebibliography}{}

\bibitem[Bartroff et~al., 2022]{Bartroff22}
Bartroff, J., Lorden, G., and Wang, L. (2022).
\newblock Optimal and fast confidence intervals for hypergeometric successes.
\newblock {\em The American Statistician}.
\newblock To appear.

\bibitem[Schilling and Doi, 2014]{Schilling14}
Schilling, M.~F. and Doi, J.~A. (2014).
\newblock A coverage probability approach to finding an optimal binomial
  confidence procedure.
\newblock {\em The American Statistician}, 68(3):133--145.

\bibitem[Schilling and Stanley, 2022]{Schilling22}
Schilling, M.~F. and Stanley, A. (2022).
\newblock A new approach to precise interval estimation for the parameters of
  the hypergeometric distribution.
\newblock {\em Communications in Statistics -- Theory and Methods},
  51(1):29--50.

\end{thebibliography}

\end{document}